\documentclass[epj,nopacs,final]{svjour}
\usepackage{graphics}
\usepackage{latexsym,overpic}
\usepackage{cite}
\usepackage{subfigure,wrapfig,multirow}
\usepackage{epsfig,color,rotating,amsmath,delarray,array}
\usepackage{makeidx,pifont,float,amssymb}
\definecolor{gray01}{gray}{0.9}
\definecolor{gray02}{gray}{0.8}
\definecolor{gray03}{gray}{0.7}
\definecolor{gray04}{gray}{0.6}
\definecolor{gray05}{gray}{0.5}
\definecolor{gray06}{gray}{0.4}
\definecolor{gray07}{gray}{0.3}
\definecolor{gray08}{gray}{0.2}
\definecolor{gray09}{gray}{0.1}

\newcommand{\er}{$\pm$}
\newcommand{\bc}{\begin{center}}
\newcommand{\ec}{\end{center}}
\newcommand{\be}{\begin{equation}}
\newcommand{\ee}{\end{equation}}
\newcommand{\ba}{\begin{eqnarray}}
\newcommand{\ea}{\end{eqnarray}}

\begin{document}

\title{Glueballs, a fulfilled promise of QCD?}
\titlerunning{Glueballs, a fulfilled promise of QCD?}
\authorrunning{E. Klempt}
\author{Eberhard Klempt}

\institute{Helmholtz-Institut f\"ur Strahlen- und Kernphysik,
Universit\"at Bonn, Germany}

\abstract{This is a contribution to the review ``50 Years of Quantum Chromdynamics"
edited by F. Gross and E. Klempt, to be published in EPJC. The contribution remembers
the early searches and explains how to find a glueball, based on its properties.  The results
of a coupled-channel analysis are presented that provides evidence for the scalar glueball
and first hints for the tensor glueball. Data on radiative decays of $\psi(2S)$ and $\Upsilon(1S)$
show scalar intensity that is likely due to glueball production.
}
\date{Received: \today / Revised version:}

\mail{klempt@hiskp.uni-bonn.de}

\maketitle

\section{Introduction}
At the {\it Workshop on QCD: 20 Years Later} held 1992 in Aachen, Heusch \cite{Heusch:1992yg}
reported on searches for glueballs, gluonium, or glue states as Fritzsch and
Gell-Mann \cite{Fritzsch:1972jv,Fritzsch:1973pi} had called this new form
of matter. Glueballs are colorless bound states of gluons and should exist when their
newly proposed quark-gluon field theory yields a correct description of  the
strong interaction. The title of Heusch's talk
{\it Gluonium: An unfulfilled promise of QCD?} expressed the disappointment of a
glueball hunter: At that time there was some - rather weak - evidence for glueball candidates but
there was no convincing case.
In 1973,  the $e^+e^-$ storage ring SPEAR at the Stanford Linear Accelerator Center had
come into operation and one year later, the $J/\psi$ resonance was
discovered~\cite{SLAC-SP-017:1974ind} - this was the very first SPEAR publication
on physics. The $J/\psi$ resonance and its radiative decay became and still is the prime reaction
for glueball searches.

One of the first glueball candidates was the
$\iota(1440)$ \cite{Scharre:1980zh,Edwards:1982nc}. The name $\iota$ stood for
the ``number one" of all glueballs to be discovered.
It was observed as very strong signal with pseudoscalar quantum numbers
in the reaction $J/\psi\to \gamma K\bar K\pi$. Its \,mass \,was \,not \,too \,far \,from \,the
\,bag-model \,pre\-diction (1290\,MeV)~\cite{Jaffe:1975fd}.
Now the $\iota(1440)$ is supposed to be split into two states,
$\eta(1405)$ and $\eta(1475)$, where the lower-mass meson is still discussed as glueball
candidate even though its mass is incompatible with lattice gauge calculations. They find
the mass of the pseudoscalar glueball above 2\,GeV.

A second candidate was a resonance called $\Theta(1640)$ \cite{Edwards:1981ex,Dunwoodie:1997an}.
It was seen in the reaction $J/\psi\to \gamma\eta\eta$
and confirmed - as $G(1590)$ - by the GAMS collaboration in $\pi^- p\to \eta\eta n$
\cite{Serpukhov-Brussels-AnnecyLAPP:1983xdr}. Its quantum numbers shifted from $J^{PC} = 2^{++}$ to $0^{++}$,
and its mass changed to 1710\,MeV. This resonance still plays an important role in the glueball
discussion.

A third candidate, or better three candidates, were  observed in the OZI rule violating
process $\pi^-p\to \phi\phi n$ \cite{Etkin:1982bw,Etkin:1987rj}. Three $\phi\phi$
resonances at 2050, 2300 and 2350 MeV were reported. I remember Armenteros
saying: {\it When you have found one glueball, you have made a discovery. When you
find three, you have a problem.} Now I believe that this was a very early
manifestation of the tensor glueball.

The situation was not that easy at that time as described here. Nearly for each observation, there were
contradicting facts, and Heusch concluded his talk at the QCD workshop
with the statement: {\it there is no smoking-gun candidate for gluonium $\cdots$.}
At this workshop, I had the honor to present the results
of the Crystal Barrel experiment at LEAR and to report the discovery of two new scalar mesons,
$f_0(1370)$ and $f_0(1500)$, and I was convinced,  Heusch was wrong:
 $f_0(1500)$ was the glueball! And I turned down my internal critical voice which told me that
in my understanding of $\bar pN$ annihilation, this process is not particularly suited
to produce glueballs \cite{Klempt:1993dw,Klempt:2005pp}. Our glueball $f_0(1500)$
was not seen in radiative $J/\psi $ decays where a glueball should stick out
like a tower in the landscape. The $f_0(1500)$ as scalar glueball? That could not be the full truth!

\section{QCD predictions}
\subsection{Glueball masses}
First estimates of the masses of glueballs were based on bag models. The color-carrying gluon
fields were required to vanish on the surface of the bag. Transverse electric and transverse magnetic
gluons were introduced populating the bag. The lowest excitation modes were predicted to
have quantum numbers $J^{PC}=0^{++}$ and $2^{++}$ and to be degenerate in mass
with $M=960$\,MeV \cite{Jaffe:1975fd,Johnson:1975zp}. A very early review  can be
found in Ref.~\cite{Robson:1977pm}.

\begin{table*}[hbt]
\begin{center}
\renewcommand{\arraystretch}{1.2}
\caption{\label{Predictions-EK}Masses of low-mass glueballs, in units of MeV. Lattice QCD results are  taken from
Refs.~\cite{Chen:2005mg,Athenodorou:2020ani} (quenched) and Ref.~\cite{Gregory:2012hu} (unquenched).
Szczepaniak and Swanson~\cite{Szczepaniak:2003mr} construct of a quasiparticle gluon
basis for a QCD Hamiltonian. Results from QCD sum rule results are given in Ref.~\cite{Chen:2021bck},
from using Dyson-Schwinger equations in \cite{Huber:2021yfy,Huber:2020ngt}, and from a graviton-soft-wall model  in Ref.~\cite{Rinaldi:2021dxh}.}
\begin{tabular}{c|c|c|c|c|c|c|c}
\hline\hline
Glueball & Ref.~\cite{Chen:2005mg} & Ref.~\cite{Athenodorou:2020ani}
 &Ref.~\cite{Gregory:2012hu}& ~~~Ref.~\cite{Szczepaniak:2003mr}~~~& Ref.~\cite{Chen:2021bck}&Ref.~\cite{Huber:2021yfy}&~~~Ref.~\cite{Rinaldi:2021dxh}~~~
\\ \hline \hline
$|0^{++}\rangle$  & $1710 \pm 50 \pm 80$                & ~~$1653 \pm 26$~~
 & ~~$1795 \pm 60$~~ & $1980$          & $1780^{+140}_{-170}$&$1850\pm130$&1920
\\
$|2^{++}\rangle$   & $2390 \pm 30 \pm 120$                & $2376 \pm 32$
 & $2620 \pm 50$ & $2420$              & $1860^{+140}_{-170}$&$2610\pm180$&2371
\\
$|0^{-+}\rangle$   & $2560 \pm 40 \pm 120$                & $2561 \pm 40$
  & --& $2220$                & ~~$2170\pm110$~~&$2580\pm180$ &
\\ \hline \hline
\end{tabular}
\label{tab:comparison}
\end{center}
\end{table*}

The bag model is obsolete nowadays. Most reliable are presumably simulation of QCD
an a lattice (see Ref.~\cite{Rothe:2012nt} for an introduction). In lattice gauge theory, the spacetime is
rotated into an Euclidean space by the transformation
$t\to i\,t$ and then discretized into a lattice with sites separated by a distance in space and time.
The gauge fields are defined as links between neighboring lattice points, closed loops of
the link variables (Wilson loops) allow for the calculation of the action density. Technically,
gluons on a space-time lattice struggle against large vacuum fluctuations of the correlation
functions of their operators, the signal-to-noise ratio falls extremely rapidly as the separation
between the source and sink is increased. These difficulties can be overcome by anisotropic
space-times with coarser space and narrow time intervals~\cite{Chen:2005mg,Morningstar:1999rf}.
Fermion fields are defined at lattice sites. Different techniques have been developed to include
fermions in lattice calculations~\cite{Athenodorou:2020ani}. The effect of see quarks on glueball masses seems to be
small~\cite{Gregory:2012hu}.

Recently, a number of different approaches were chosen to approximate QCD by a model that
is solvable analytically.
Szczepaniak and Swanson~\cite{Szczepaniak:2003mr} constructed a quasi-particle gluon
basis for a QCD Hamiltonian in Coulomb gauge that was solved analytically. A full glueball spectrum
was calculated with no free parameter.
The authors of Ref.~\cite{Chen:2021bck} constructed relativistic two- and three-gluon glueball
currents and applied them to perform QCD sum rule analyses of the glueball spectrum.
The Gie\ss en group calculated masses of ground and excited glueball states
using a Yang-Mills theory and a functional approach based on a
truncation of Dyson-Schwinger equations and
a set of Bethe-Salpeter equations derived from a three-particle-irreducible effective action
\cite{Huber:2021yfy,Huber:2020ngt}.

AdS/QCD relies on a correspondence between a five dimensional classical theory with an AdS metric
and a supersymmetric conformal quantum field theory in four dimensions. In the bottom-up approach, models with
appropiate operators are constructed in the classical AdS theory with the aim of
resembling QCD as much as possible. Confinement is generated by a hard wall cutting off AdS space
in the infrared region, or spacetime is capped off smoothly by a soft wall to break the conformal
invariance. Rinaldi and Vento~\cite{Rinaldi:2021dxh} calculated the glueball mass spectrum within
AdS/QCD. The results on glueball masses are summarized in Table~\ref{Predictions-EK}.

\subsection{The width of glueballs}
Glueballs are often assumed to be narrow. $\phi$ decays into $\rho\pi$ are suppressed
since the primary $s\bar s$ pair needs to annihilate and a new $q\bar q$ pair needs to be
created. In glueball decays, there is no pair to be annihilated but a $q\bar q$ pair needs to be
created. If the OZI rule suppresses the decay by a factor 10 to 100, we might expect the width
of glueballs to be suppressed by a factor 3 to 10. Assuming a ``normal" width of 150\,MeV,
a glueball at 1600\,MeV could have a width of 15 to 50\,MeV.  This argument is supported
by arguments based on the $1/N_c$ expansion (see, e.g., Ref.~\cite{RuizdeElvira:2010cs}).

Narison applied QCD sum rules~\cite{Narison:1996fm}. Assuming a mass of 1600\,MeV, he
 calculated the $4\pi$ width of the scalar glueball to 60 to 138\,MeV
while the partial decay width of the tensor glueball at 2\,GeV to pseudoscalar mesons should be
less than 155\,MeV. Calculations on the lattice gave a partial decay width for decays
into pseudoscalar mesons of 108\er29 MeV for a scalar glueball mass of 1700\,MeV~\cite{Sexton:1996ed}.
In a semi-phenomenological model, Burakovsky and Page find that the width of the scalar
glueball (at 1700\,MeV) should exceed 250 to 390\,MeV. A flux tube model predicted
the mass of the glueball of lowest mass to 1680\,MeV and its width to 300\,MeV~\cite{Iwasaki:2003cr}.
In a field theoretical approach
with an effective Coulomb gauge the glueball width was estimated to 100\,MeV~\cite{Bicudo:2006sd}.

\subsection{Radiative yields}
The study of radiative decays of the $J/\psi$ meson is the prime path to search for
glueballs with masses of less than $\sim$2500 MeV.

Gui \textit{et al.}~\cite{Gui:2012gx} calculated the yield of a scalar glueball having a mass of 1710\,MeV on
lattice and found
\ba
BR_{J/\psi\to\gamma G_{0^{++}}}(TH)&=& (3.8\pm0.9)\cdot 10^{-3}.
\ea
For higher glueball masses the yield increases.

Narison gave a mass dependent formula derived from sum rules. For a mass of 1865\,MeV, a
yield of about $10^{-3}$ is predicted \cite{Narison:1996fm}.

The tensor glueball is expected~\cite{Chen:2014iua} to be observed with a branching ratio
\ba
BR_{J/\psi\to\gamma G_{2^{++}}}(TH)&=& (11\pm2)\cdot 10^{-3}.
\ea
Production of the pseudoscalar glueball seems to be considerably smaller. For
a mass of 2395 (or 2560)\,MeV, the authors of Ref.~\cite{Gui:2019dtm} find
\ba
BR_{J/\psi\to\gamma G_{0^{-+}}}(TH)&=& (0.231\pm0.080)\cdot 10^{-3}\nonumber\\
 {\rm or} &=&(0.107\pm 0.037)\cdot 10^{-3}.
\ea
These are very significant yields, and the glueballs must be found provided
they can be identified convincingly as glueballs amidst their $q\bar q$ companions.

\section{\label{sec:glueball-tag}How to identify a glueball}
\begin{figure}[b]
\centering
\includegraphics[width=0.45\textwidth]{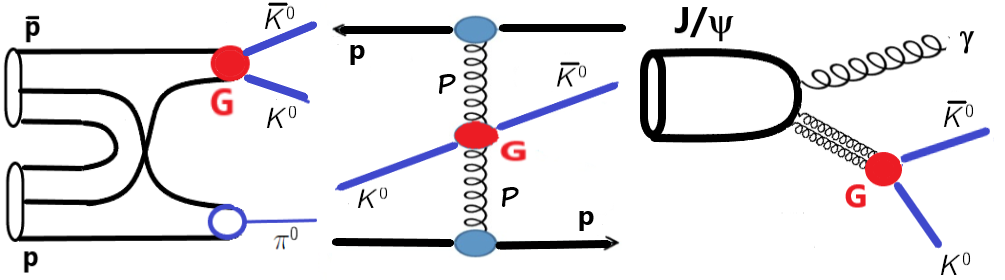}
\caption{\label{fig:reactions}Reactions most relevant for glueball searches.
Left: $\bar pp$ annihilation; middle: Pomeron-Pomeron fusion; right: radiative $J/\psi$
decays. The glueball is supposed to decay into $K^0\bar K ^0$.
}
\end{figure}

Figure~\ref{fig:reactions} shows the prime reactions in which glueballs have been searched for.

\subsection{\boldmath$N \bar N$ annihilation}
A decisive step forward in the search for glueballs was the discovery of
two scalar isoscalar states in $\bar p p$ annihilation at rest. With the large statistics
available at the Low Energy Anitiproton Ring (LEAR) at CERN,
$f_0(1370)$ and $f_0(1500)$ were identified in several final states.
A large fraction of the data taken at LEAR is still used jointly with data on radiative $J/\psi$
decays in a coupled-channel analysis. Glueballs decay via $q\bar q$ pair creation.
Hence they can be produced via $q\bar q$ annihilation. Meson production in $\bar p p$ annihilation
was studied by the ASTERIX, OBELIX and Crystal Barrel experiments at LEAR and
is a major objective of the PANDA collaboration at the GSI.
\subsection{Central production}
In central production, two hadrons (e.g. two protons) scatter in forward direction via the
exchange of Pomerons. Pomerons are supposed to be glue-rich, hence glueballs can be
formed in Pomeron-Pomeron fusion.  This process was studied extensively at CERN
by the WA76 and WA102 collaborations and is now investigated with the STAR detector
at BNL. In the  WA102 experiment, $f_0(1370)$ and $f_0(1500)$ were confirmed and
$f_0(1710)$ was added to the number of scalar resonances.
\subsection{radiative $J/\psi$ decays}
In radiative $J/\psi$ decays, the primary $c\bar c$ pair converts into two gluons and a photon.
The two gluons are mainly produced in $S$-wave, the two gluons
can form scalar and tensor glueballs which should be
produced abundantely.
The large statistics available from BESIII at Beijing makes this reaction
the most favorable one for glueball searches. Radiative decays of heavy mesons is the
only process for which glueball yields have been calculated. The data will be discussed
below in more detail.
\subsection{Decay analysis}
\begin{figure}[pb]
\centering
\includegraphics[width=0.48\textwidth]{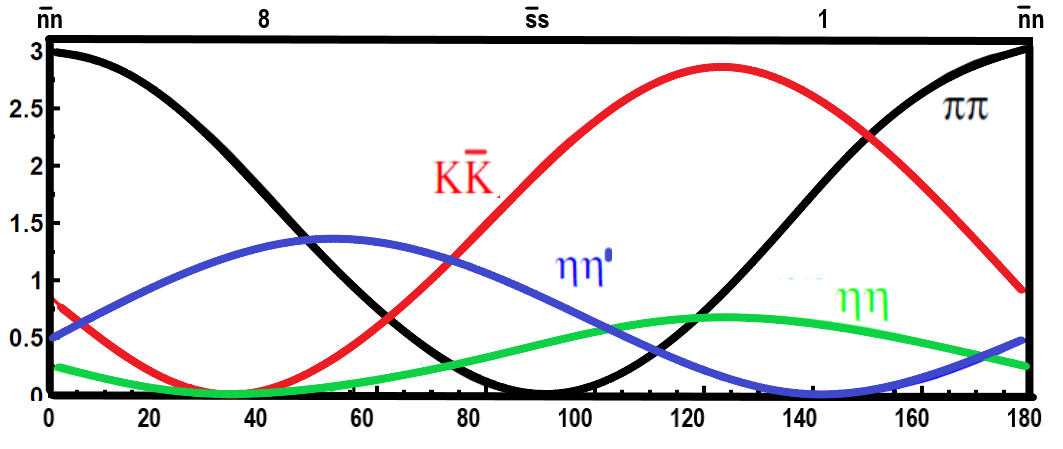}
\caption{\label{fig:mixing}Decay probabilities of mesons for decays
into two pseudoscalar mesons as a function of the scalar mixing angle~\cite{Klempt:2021wpg}.
}
\end{figure}
The decay of mesons into two pseudoscalar mesons is governed by SU(3)$_F$. In a
meson nonet, there are two isoscalar mesons, one lower in mass the other one higher,
which both contain a $n\bar n=(u\bar u+d\bar d)/\sqrt 2$ and a $s\bar s$ component and
 are mixed with the mixing angle $\varphi$.
Figure~\ref{fig:mixing} shows the SU(3)$_F$ squared matrix elements for
meson decays into two pseudoscalar mesons as a function of the scalar mixing angle.\\

\qquad$
\left( \begin{array}{c}
f^H\\
f^L\\
\end{array}\right)
=
\left( \begin{array}{cc}
\cos\varphi^{\rm s}& -\sin\varphi^{\rm s}  \\
\sin\varphi^{\rm s}  & \cos\varphi^{\rm s}\\
\end{array}\right)
\left( \begin{array}{c}
|n\bar n> \\
|s\bar s>
\end{array}\right)$
\setcounter{equation}{6} \hfill(6)\\[-2ex]
\subsection{Supernumery}
The three scalar isoscalar mesons $f_0(1370)$, $f_0(1500)$ and $f_0(1710)$ played
an important role in the glueball discussion. Amsler and Close~\cite{Amsler:1995tu,Amsler:1995td}
suggested to interpret
these three states as the result of mixing of the two expected isoscalar states
with the scalar glueball.
\be
\left( \begin{array}{r}
f_0(1370)  \\
f_0(1500)\\
f_0(1710) \\
\end{array}
\right)
=
\left( \begin{array}{rrr}
x_{11} & x_{12} & x_{13} \\
x_{21} & x_{22} & x_{23} \\
x_{31} & x_{32} & x_{33} \\
\end{array}
\right)
\left( \begin{array}{r}
|n\bar n\rangle  \\
|s\bar s\rangle \\
|gg\rangle \\
\end{array}
\right)\label{eq:mixEK}
\ee
These papers led to a large number of follow-up papers, references can be found
in Ref.~\cite{Klempt:2021wpg}. In all these papers, these three mesons contain
the full glueball, $\sum_j x_{ij}^2=1$ is imposed. Note that
the squared mass difference between $f_0(1370)$ and $f_0(1710)$
is slightly above 1\,GeV$^2$, the $f_0(1710)$ could also be a radial excitation (and is interpreted
as radial excitation below).
\subsection{Conclusions}
Identifying a glueball is a difficult task. The main argument in favor
of a glueball interpretation is an anomalously large production rate in $J/\psi$ decays.
It turns out that scalar mesons are organized like pseudoscalar mesons, into mainly singlet and mainly octet mesons.
A large production rate of a mainly-octet scalar isoscalar meson in radiative $J/\psi$ decays
directly points to a significant glueball content in its wave function. A second argument relies
an meson decays into pseudoscalar mesons. In presence of a glueball, a pair of mesons assigned to the same multiplet should have a decay pattern that is
incompatible with a $q\bar q$ interpretation for any mixing angle. Supernumery
is a weak argument. It requires a full understanding of the regular excitation spectrum.
Further studies are required to learn if central production is gluon-rich. The large production rates
of $f_0(1500)$, $f_0(1710)$ and $f_0(2100)$ in $\bar pp$ annihilation at collision energies above 3\,GeV
encourages glueball searches at the FAIR facility with the PANDA detector.

\section{Evidence for glueballs from coupled-channel analysis}
\begin{figure}[pt]
\centering
\includegraphics[width=0.45\textwidth]{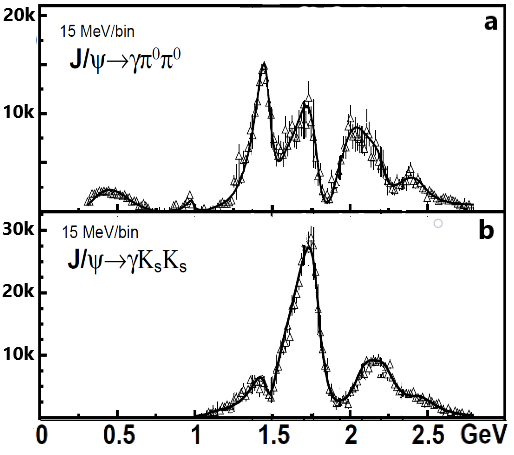}
\caption{\label{fig:pwajpsi}The squared $S$-wave tranisition amplitudes for
$J/\psi\to\pi^0\pi^0$ (a) and $J/\psi\to K^0_sK^0_s$ (b). The data points are from an
energy-independent partial-wave analysis~\cite{BESIII:2015rug,BESIII:2018ubj},
the curve represents our fit~\cite{Sarantsev:2021ein}.
}
\end{figure}

We have performed a coupled-channel partial wave analysis of radiative $J/\psi$
decays into $\pi^0\pi^0$, $K^0_sK^0_s$, $\eta\eta$,
and $\omega\phi$, constrained by the CERN-Munich data on $\pi N$ scattering, data
from the GAMS collaboration at CERN, data from BNL on $\pi\pi\to K^0_sK^0_s$,
and 15 Dalitz plots on $\bar pp$ annihilation at rest from LEAR.
Data on $K_{\rm e4}$ decays constrain the low-energy region. Fitting details
and references to the data can be found in Ref.~\cite{Sarantsev:2021ein}
\footnote{The BESIII data were fitted by Rodas {\it et al.}~\cite{Rodas:2021tyb}
with four scalar and three tensor resonances only. I have several objections against the fit.
i) It uses an amplitude in which the $J/\psi$ converts into three gluons which hadronize.
A final-state meson radiates off the photon. Since the photon is not an isospin eigenstate,
this amplitude can produce isovector mesons. This process is highly suppressed and experimentally absent.
ii) the $f_0(1370)-f_0(1500)$ interference region is not well described, neither in the mass
distribution nor in the $S-D$ phase difference. iii) The fit is neither constrained by the $\pi\pi$
$S$-wave from the CERN-Munich data nor by the data on $K_{\rm e4}$ decays. A fit with
the seven resonances used in Ref.~\cite{Rodas:2021tyb} without an isospin breaking amplitude leads to
a $\pi\pi$ $S$-wave that is extremely incompatible with the known $\pi\pi$ $S$-wave (A.V. Sarantsev,
private communication, October 2021.)
}
Figure~\ref{fig:pwajpsi} shows the data on radiative $J/\psi$
decays into $\pi^0\pi^0$, $K^0_sK^0_s$ and the fit. Ten scalar isoscalar
resonances were included in the fit. Oller~\cite{Oller:2003vf} has shown
that $f_0(500)$ is singlet-like, the $f_0(980)$ octet-like
(see also \cite{Klempt:2021nuf}). The $f_0(1500)$ is seen in Figure~\ref{fig:pwajpsi}
as a dip. This pattern was reproduced in Ref.~\cite{Sarantsev:2021ein} assuming
that $f_0(1370)$ is a singlet state and $f_0(1500)$ an octet state. Hence we assumed
that the ten mesons can be divided into two series of states, mainly-singlet states
with lower masses and mainly-octet states with higher masses.

\begin{figure}[ht]
\centering
\hspace{-2mm}\includegraphics[width=0.49\textwidth]{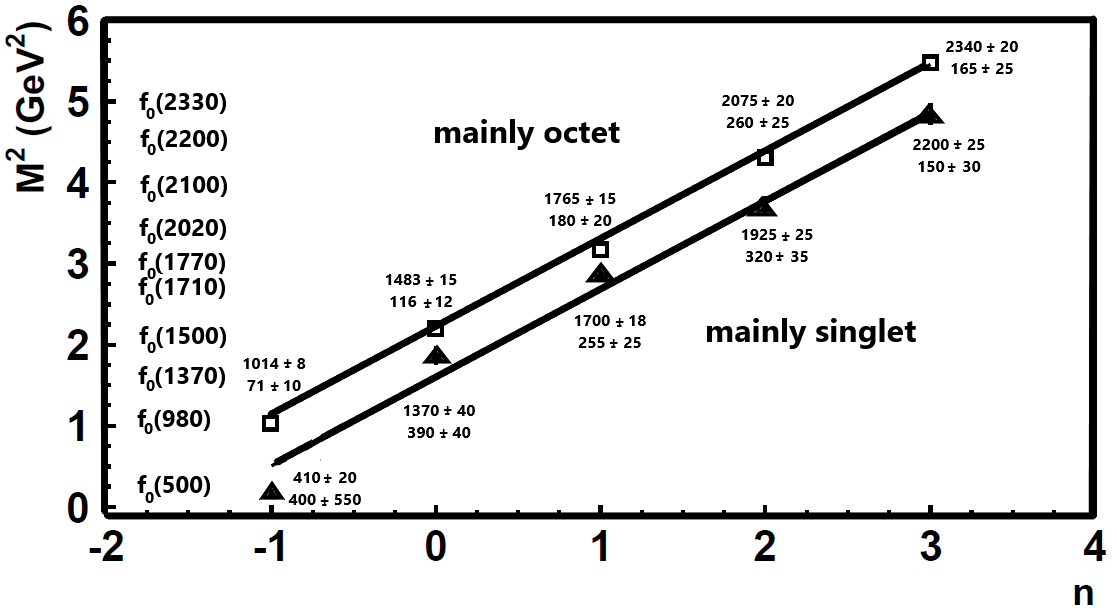}
\caption{\label{fig:scaltraj}$M^2,n$ trajectories for mainly-singlet and mainly-octet
scalar isoscalar resonances. The red dot at high masses represents a scalar state from $J/\psi\to \gamma\eta'\eta'$~\cite{BESIII:2022zel}.
Adapted from Ref.~\cite{Sarantsev:2021ein}.}
\end{figure}

In a ($M^2,n$) plot, the masses of singlet and octet states follow two
linear trajectories (see Fig.~\ref{fig:scaltraj}). Remarkably, the slope (1.1\,GeV$^{-2}$) is
close to the slope of standard Regge trajectories. The separation between the two
trajectories is given by the mass square difference between $\eta'$ and $\eta$-meson
as suggested by instanton-induced interactions~\cite{Klempt:1995ku}.
The figure includes a meson reported by the
BESIII collaboration studying $J/\psi\to \gamma\eta'\eta'$~\cite{BESIII:2022zel}. As
$\eta'\eta'$ resonance, $f_0(2480)$ is very
likely a SU(3) singlet state. Indeed, its mass
is compatible with the ``mainly-singlet"
trajectory. The figure gives the pole positions
of the eleven resonances as small inserts.
\begin{figure}[pt]
\centering
\includegraphics[width=0.40\textwidth,height=0.26\textwidth]{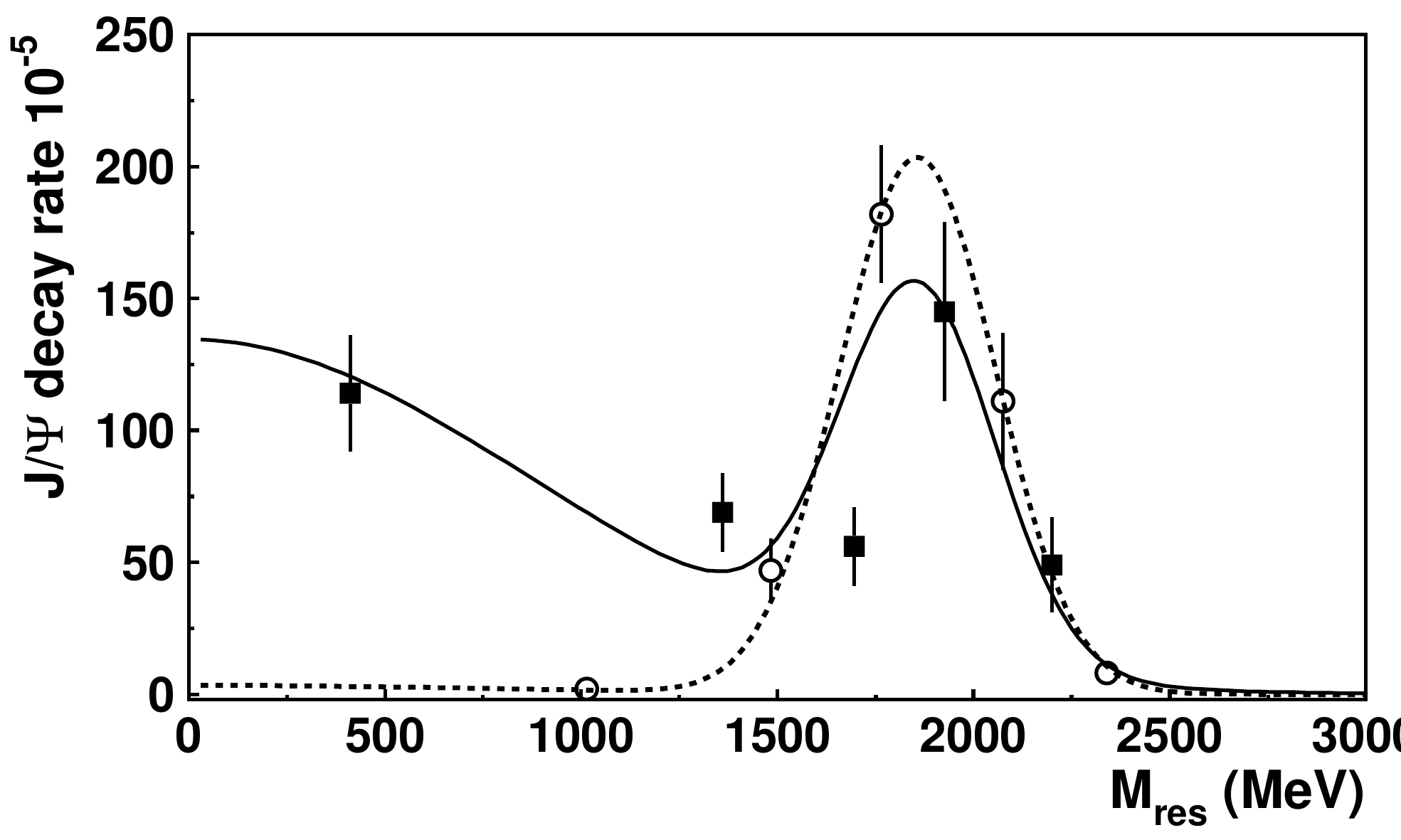}
\caption{\label{fig:yieldEK}Yield of scalar isoscalar mesons in radiative $J/\psi$ decays
into mainly-octet (open circles) and mainly-singlet mesons (full squares) as a function of their mass~\cite{Sarantsev:2021ein}.
}
\end{figure}

The total yields of scalar mesons in radiative $J/\psi$ decays - including decay modes not reported by the BES\-III collaboration - was
determined from the coupled-channel analysis~\cite{Sarantsev:2021ein} that included also
other data. The yield of mainly-octet and mainly-singlet mesons as a function of their mass
is shown in Fig.~\ref{fig:yieldEK}. Mainly-octet mesons should not be produced (or at most weakly) in radiative $J/\psi$ decays. However, they are produced abundantly, in a limited mass range
centered at about 1865\,MeV. Mainly-singlet mesons are produced over the full mass range
but show a peak structure at the same mass. This enhancement must be due to the scalar glueball
mixing into the wave functions of scalar mainly-octet and mainly-singlet mesons. A Breit-Wigner fit to
these distributions gives mass and width
\be
M_G=(1865\pm 25^{\,+10}_{\,-30})\,{\rm MeV}\quad \Gamma_G= (370\pm
50^{\,+30}_{\,-20})\,{\rm MeV}\,,\label{eq:massEK}
\ee
\phantom{z}\\[-2ex]
and the (observed) yield is determined to
\be
Y_{J/\psi\to\gamma G}=(5.8\pm 1.0)\,10^{-3}\,.\label{eq:widthEK}
\ee

\section{Meson-glueball mixing}
Earlier attempts to identify the glueball have in common that
the full glueball is distributed among the three states $f_0(1370)$, $f_0(1500)$ and $f_0(1710)$.
Inspecting Fig.~\ref{fig:pwajpsi},
this seems not to be obvious: Above 1\,GeV, four peaks with three valleys are seen, and
there is no reason why one particular region should be more gluish than the other ones.
The yield of scalar mesons sees the glueball contribution distributed over several resonances.

We did not impose that the full glueball should be seen in these three states
nor that we must see the full glueball at all. We fitted the decay modes
of pairs of scalar mesons, one mainly-singlet one mainly-octet, and allowed for a glueball component~\cite{Klempt:2021wpg}.
\ba
~\hspace{-2mm}f^{\rm nH} _0(xxx)=\left(n\bar n\cos\varphi^{\rm s} _{\rm n}-s\bar s\sin\varphi^{\rm s} _{\rm n}\right )\cos\phi^G _{\rm nH} + G\sin\phi^G _{\rm nH}\nonumber\\
~\hspace{-4mm}f^{\rm nL} _0(xxx)=\left(n\bar n\sin\varphi^{\rm s} _{\rm n}+s\bar
s\cos\varphi^{\rm s} _{\rm n}\right )\cos\phi^G _{\rm nL} + G\sin\phi^G _{\rm nL} \nonumber\\[-0.5ex]
\ea
$\varphi^{\rm s} _{\rm n}$ is the scalar mixing angle, $\phi^G _{\rm nH}$ and $\phi^G _{\rm nL}$
are the meson-glueball mixing angles of the high-mass state H and of the low-mass state L in the nth nonet.
The fractional glueball content of a meson is given by $\sin^2\phi^G _{\rm nH}$ or $\sin^2\phi^G _{\rm nL}$.

\begin{figure}[t]
\centering
\includegraphics[width=0.40\textwidth,height=0.26\textwidth]{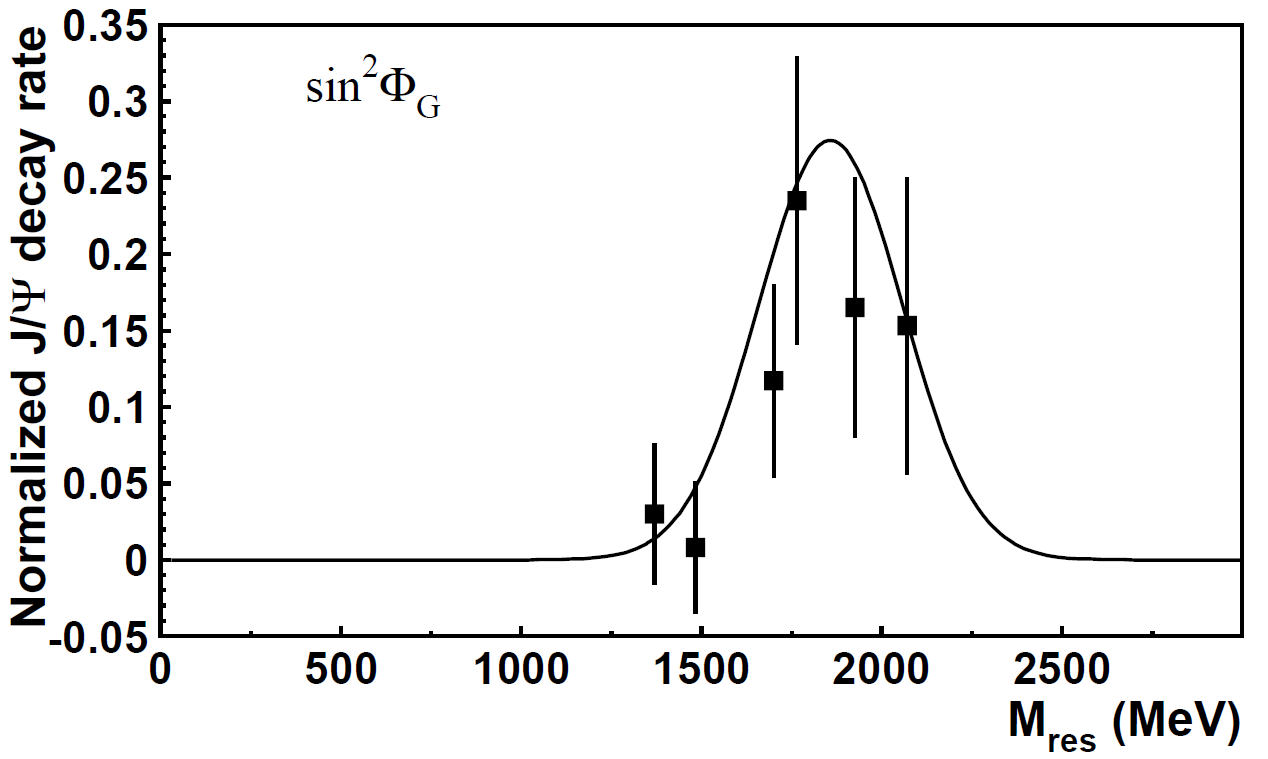}
\caption{\label{fig:gluecontEK}The glueball content of scalar mesons.
Black squares: $\sin^2\varphi^{\rm s} _{\rm n}$, solid curve: Breit-Wigner resonance
with area 1~\cite{Klempt:2021wpg}. 
}
\end{figure}

With this mixing scheme and the SU(3) coupling constant (see Fig.~\ref{fig:mixing}),
we have fitted the meson decay modes and have thus determined the glueball content of
the eight high-mass scalar mesons. Figure~\ref{fig:gluecontEK} shows the glueball
fraction in the scalar mesons.

The glueball fractions derived from the decay analysis of pairs of scalar mesons add up to
a sum that is compatible with 1. The distribution of the glueball fraction in
Fig.~\ref{fig:gluecontEK} is identical to the distribution of yields in Fig.~\ref{fig:yieldEK}.
This is a remarkable confirmation that the scalar glueball of lowest mass has been identified
and has mass and width as given in Eqn.~(\ref{eq:massEK}) and a
yield as given in Eqn.~(\ref{eq:widthEK}).

\section{Comparison with LHCb data}

\begin{figure}[b]
\centering
\includegraphics[width=0.40\textwidth]{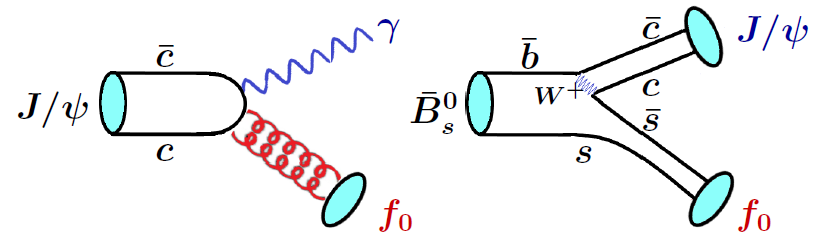}
\caption{\label{Reactions}In radiative $J/\psi$ decays two gluons, in $\bar B^0_s\to J/\psi +s\bar s$, a
$s\bar s$ pair may convert into a scalar meson.
}
\end{figure}

Most striking is the mountain landscape
above 1500\,MeV in the data on radiative $J/\psi$ decays. In these decays
a $c\bar c$ pair converts into gluons which hadronize (see Fig.~\ref{Reactions}, left).
The huge peak in the $K\bar K$
mass spectrum at 1750\,MeV and the smaller one at 2100\,MeV decay are produced with
two gluons in the initial state.
This is to be contrasted with data on $B_s ^0$ and $\bar B_s ^0$ decays into $J/\psi+\pi^+\pi^-$~\cite{LHCb:2014ooi} and
$K\bar K$~\cite{LHCb:2017hbp}. In this reaction, a primary $s\bar s$ pair -- recoiling against the $J/\psi$ -- converts
into the final state mesons  (see Fig.~\ref{Reactions}, right). We have included the spherical harmonic moments in the coupled channel
analysis that describes the radiative $J/\psi$ decays~\cite{Sarantsev:2022tdi}.
High-mass scalar mesons are only weakly produced in $B_s ^0$
decays with $s\bar s$ in the initial state. The strong peak in the $K\bar K$ invariant mass at 1750\,MeV in Fig.~\ref{fig:pwajpsi}
is nearly absent in $B_s ^0\to J/\psi\,K\bar K$\,!

\begin{figure}[t]
\centering
\includegraphics[width=0.40\textwidth]{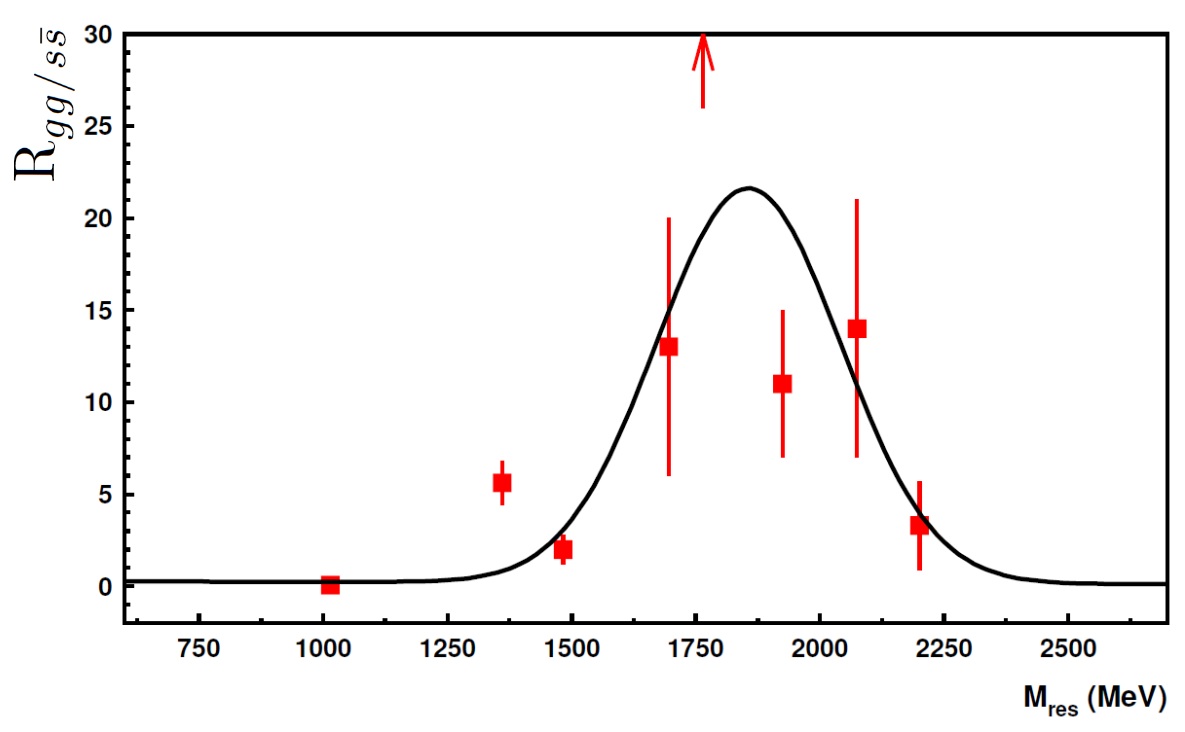}
\caption{\label{fig:glueproduction}The ratio R$_{gg/s\bar s}$ of the frequencies for
$J/\psi\to\gamma\,f_0$  and $B_s ^0\to J/\psi\,f_0$ with $f_0$ decaying into $\pi\pi$ or $K\bar K$.
}
\end{figure}

Figure~\ref{fig:glueproduction} shows the ratio of the
decay frequencies of $J/\psi\to\gamma\,f_0$  and $B_s ^0\to J/\psi\,f_0$ with $f_0$ decaying into $\pi\pi$ or $K\bar K$.
The $f_0(980)$ is likely a mainly-octet state, little produced in radiative $J/\psi$ decays but strongly with $s\bar s$
in the initial state. On the contrary, $f_0(1770)$ is seen as strong peak in radiative $J/\psi$ but very weakly only
in $B_s ^0$ decays. The uncertainties are large, but the ratio of the decay frequencies is fully compatible with
the shape of the glueball derived above.

This is highly remarkable: the two
gluons in the initial state must be responsible for the production
of resonances that decay strongly into $K\bar K$
but are nearly absent when $s\bar s$ pairs are in the initial state. Also, there is a rich structure in the $\pi\pi$ mass spectrum produced in radiative $J/\psi$ decays but little activity only when the initial state is an
$s\bar s$ pair: The
rich structure stems from gluon-gluon dynamics. Similar conclusions can be drawn~\cite{Klempt:2021nuf} from a comparison of the
invariant mass distributions from radiative $J/\psi$ decays with the pion and kaon form factors~\cite{Ropertz:2018stk}.
Their square is proportional to the cross sections.
The $f_0(980)$ resonance dominates both formfactors
but is nearly absent in radiative $J/\psi$ decays: The  $f_0(980)$ has large $n\bar n$ and $s\bar s$ components mixed to a dominant SU(3) octet component. The large intensity above 1500\,MeV in radiative $J/\psi$ decays is absent
when not two gluons but an $s\bar s$ pair is in the initial state: the mountainous structure in radiative $J/\psi$ decays
is produced by gluons and not by $q\bar q$ pairs: The structure is due to the scalar glueball.

\section{A trace of the  tensor glueball}
The tensor glueball is predicted with an even higher yield~\cite{Chen:2014iua}:
\ba
\Gamma_{J/\psi\to\gamma/ G_{2^{++}}}/\Gamma_{\rm tot} &=&(11\pm2) 10^{-3}\,.\label{tensor-pred}
\ea
The yield of $f_2(1270)$ in radiative $J/\psi$ decays is $(1.64\pm 0.12)10^{-3}$,
about six times weaker than the predicted rate for the tensor glueball!
\begin{figure}[t]
\begin{center}
\vspace{-4mm}
\begin{tabular}{cc}
\hspace{-4mm}\begin{overpic}[width=0.27\textwidth,height=0.32\textwidth]{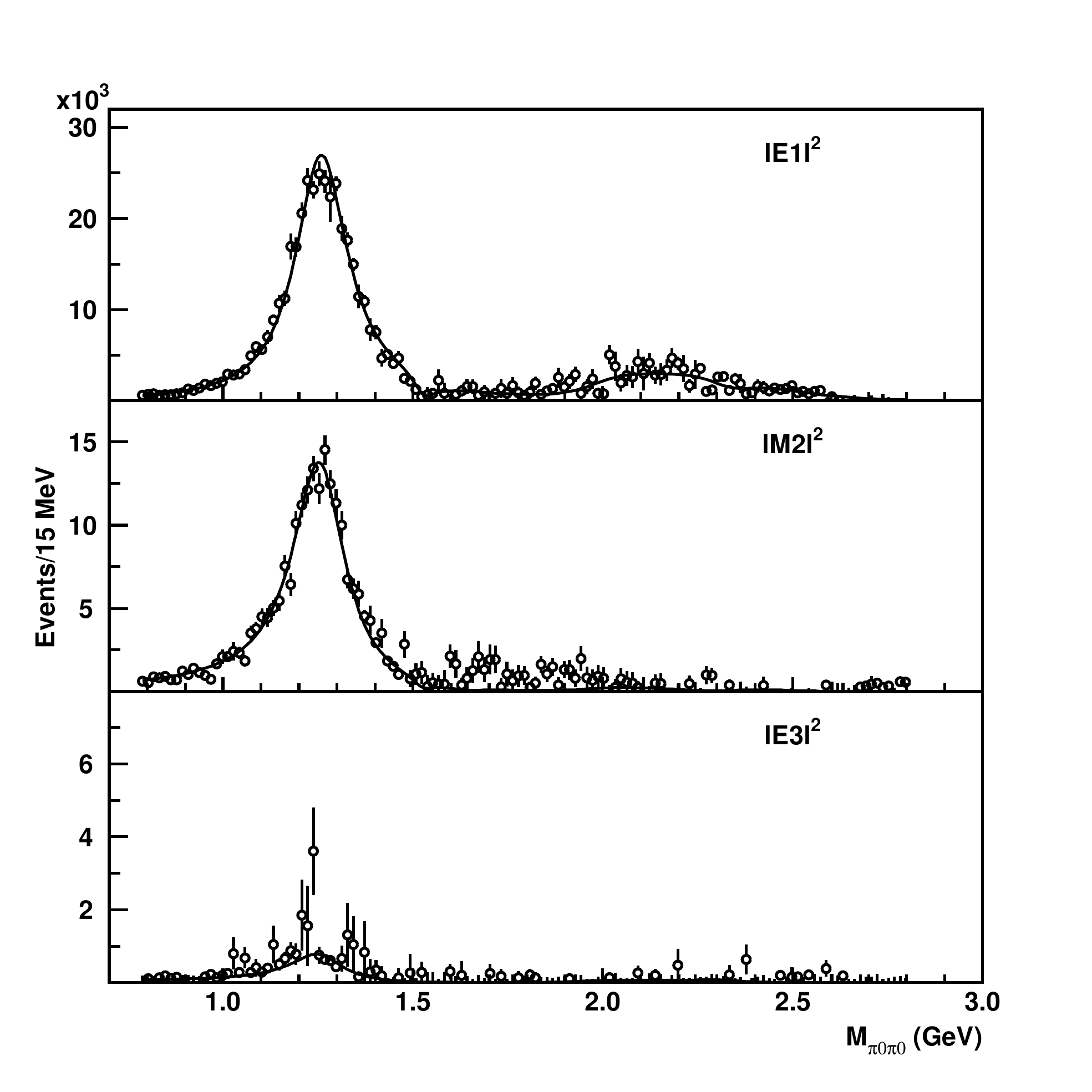}
\put(11,84){\bf\large a}
\put(11,57){\bf\large b}
\put(11,30){\bf\large c}
\end{overpic}&\hspace{-10mm}
\begin{overpic}[width=0.27\textwidth,height=0.32\textwidth]{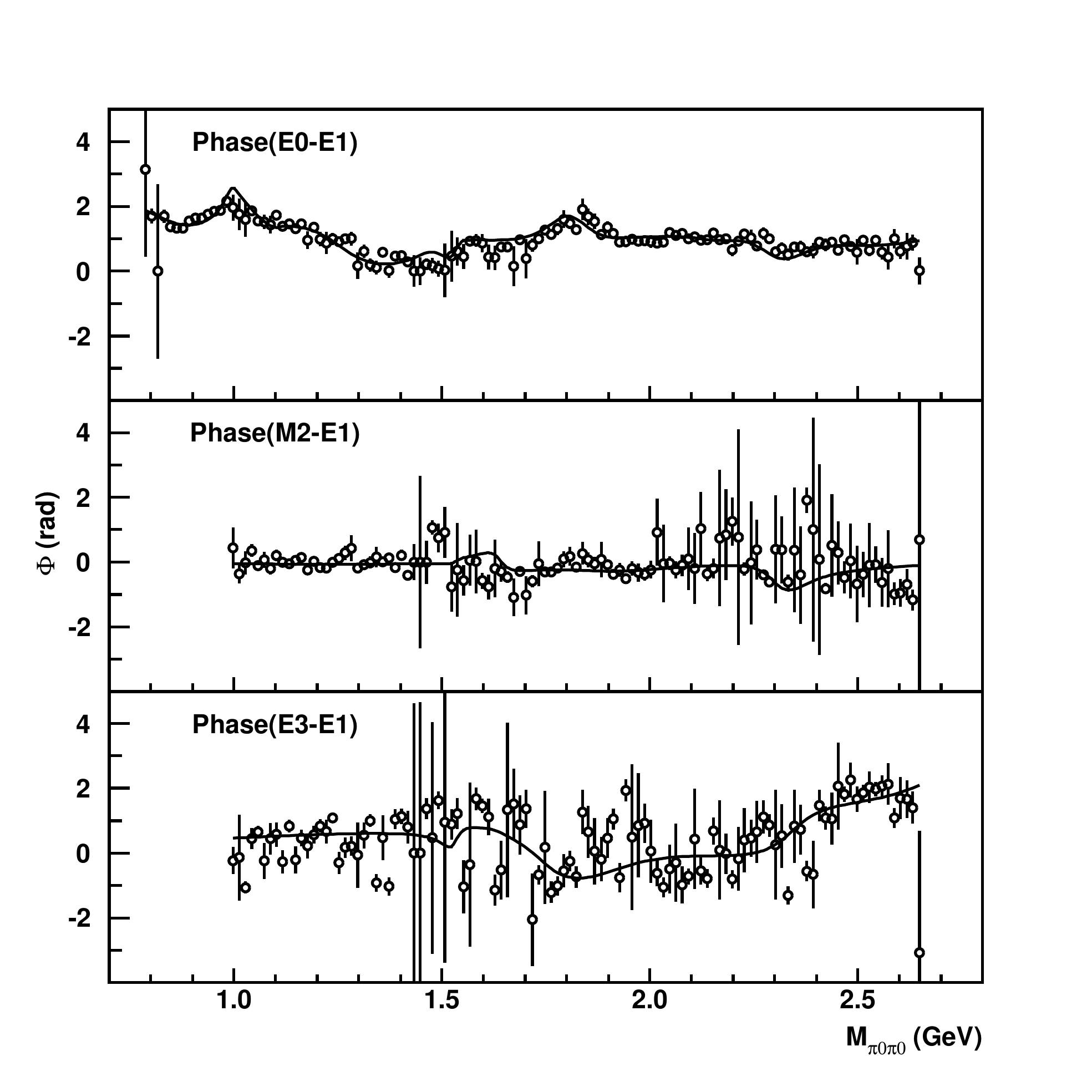}
\put(11,84){\bf\large d}
\put(11,57){\bf\large e}
\put(11,30){\bf\large f}
\end{overpic}\\[-4ex]
\hspace{-4mm}\begin{overpic}[width=0.27\textwidth,height=0.32\textwidth]{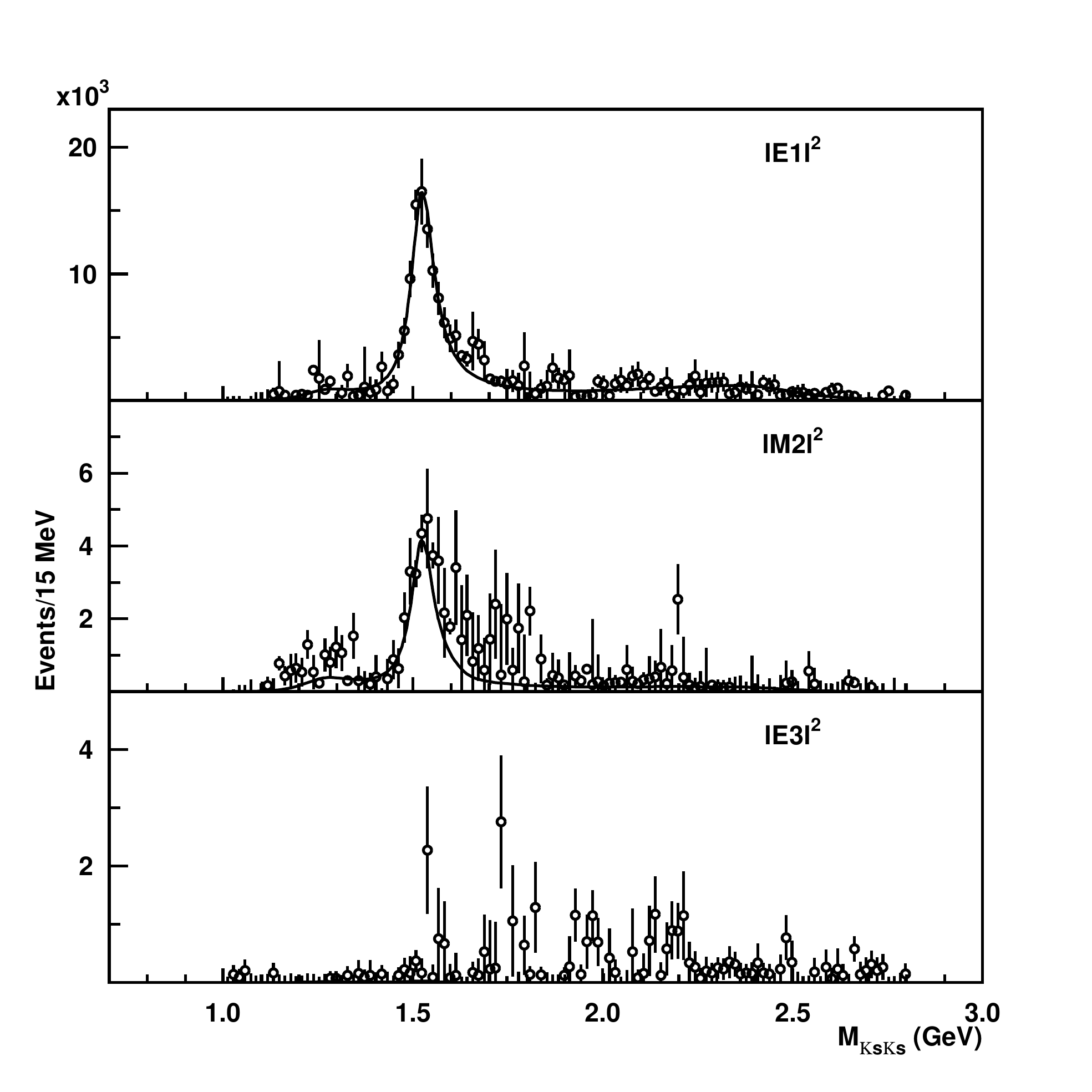}
\put(11,84){\bf\large a}
\put(11,57){\bf\large b}
\put(11,30){\bf\large c}
\end{overpic}&\hspace{-10mm}
\begin{overpic}[width=0.27\textwidth,height=0.32\textwidth]{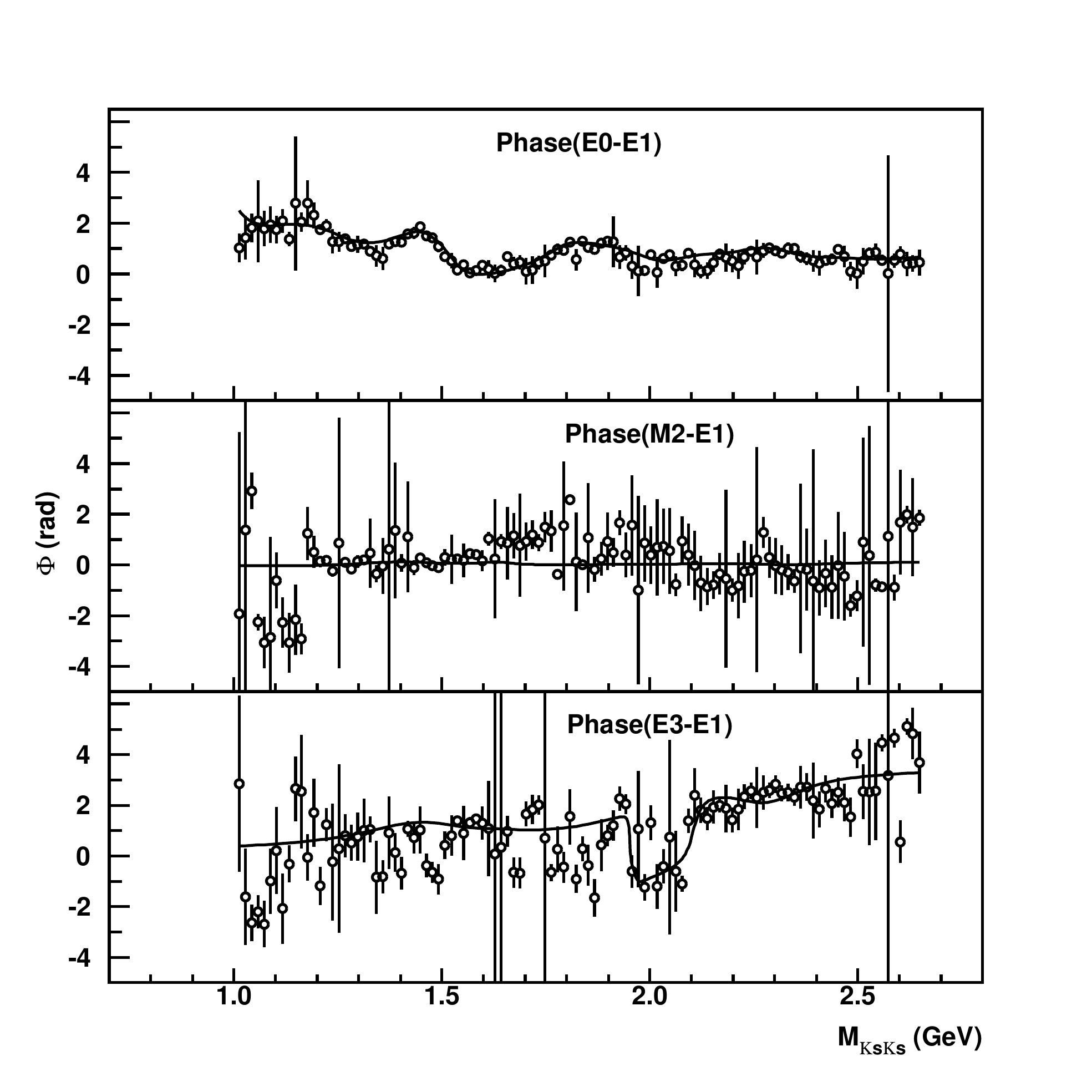}
\put(11,84){\bf\large d}
\put(11,57){\bf\large e}
\put(11,30){\bf\large f}
\end{overpic}
\end{tabular}
\end{center}
\caption{\label{jpsia}$D$-wave intensities and phases for radiative $J/\psi$ decays into
$\pi^0\pi^0$ (top subfigures) and $K_s\,K_s$ (bottom subfigures)  from
Ref.~\cite{BESIII:2015rug,BESIII:2018ubj}. The subfigures show
the $E1$ (a), $M2$ (b) and $E3$ (c) squared amplitudes
and the phase differences between the $E0$ and $E1$ (d) amplitudes, the $M2$ and $E1$ (e) amplitudes,
and the $E3$ and $E1$ (f) amplitudes as functions of the meson-meson invariant mass. The
phase of the $E0$ amplitude is set to zero. The  curve represents our best fit.
}
\end{figure}
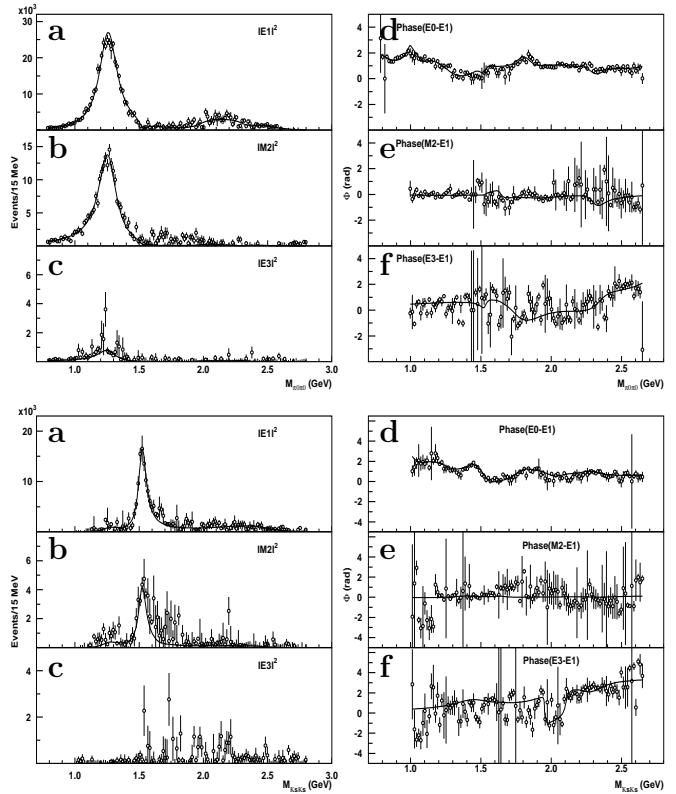
 Bose symmetry implies that the $\pi^0\pi^0$ or $K_sK_s$ pairs are
limited to even angular momenta, practically, only $S$ and $D$-waves contribute.
The scalar intensity originates from the electric dipole transition $E0$.
Three electromagnetic amplitudes $E1, M2$, and $E3$ excite tensor mesons.
Figure~\ref{jpsia} shows these three amplitudes and the relative phases.

Two fits were performed \cite{Klempt:2022qjf}. One fit describes the mass distribution only.
Apart from the well known $f_2(1270)$ and $f_2'(1525)$ the fit needs
one high-mass resonance with
\be
M=(2210\pm 60)\,{\rm MeV}; \ \ \Gamma=(360\pm 120)\,{\rm
MeV}\,,
\ee
where the error includes systematic studies with or without additional low-yield resonances.
The enhancement was called  $X_{2}(2210)$.  In this fit, the phases are not well described.  Figure \ref{jpsia} shows a fit in which
the 2200\,MeV region is described by  three tensor resonances
with masses and widths of about
$(M,\Gamma) \,= \,(2010, 200)$,  $(2300, 150)$,  and
$(2340, 320)$ MeV. These resonances had been observed by Etkin {\it et al.}~\cite{Etkin:1987rj}
in the reaction $\pi^-p\to \phi\phi n$. The unusual production characteristics were interpreted in Ref.~\cite{Etkin:1987rj}
as evidence that {\it these states are produced by $1 - 3$ glueballs.}

The total observed yield of  $X_{2}(2210)$ in $\pi\pi$ and $K\bar K$ is $(0.35\pm 0.15)\,10^{-3}$,
far below the expected glueball yield. We assume the glueball is -- like the scalar glueball  --
distruted over several tensor mesons. Adding up all contributions from
tensor states above 1900\,MeV seen in radiative $J/\psi$ decays, one obtains
\be
\sum_{M=1.9\,{\rm GeV}}^{M=2.5\,{\rm GeV}} Y_{J/\psi \to \gamma f_2} = (3.1\pm 0.6)\,10^{-3}\,,
\ee
which is a large yield even though still below the predicted yield.

\section{How to find the pseudoscalar glueball}
\begin{figure}[t]
\begin{center}
\begin{tabular}{cc}
\hspace{-4mm}\includegraphics[width=0.25\textwidth,height=0.2\textwidth]{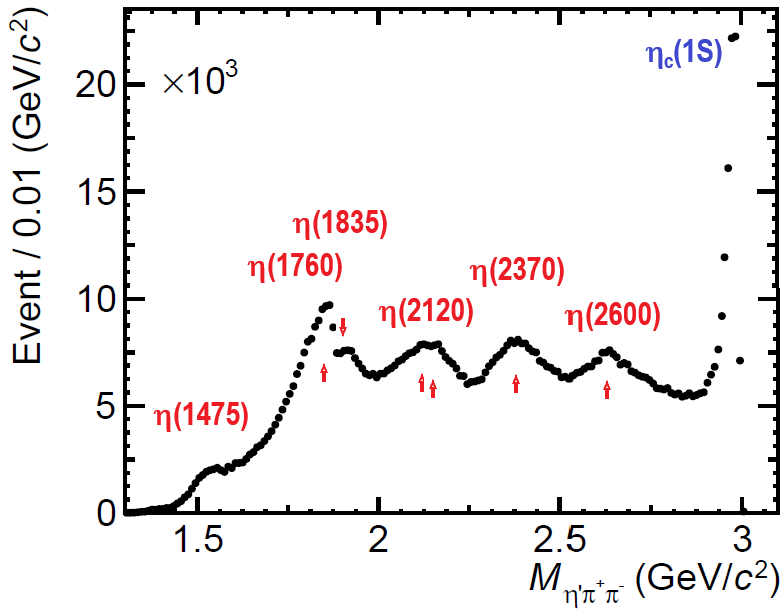}&\raisebox{3mm}{
\hspace{-4mm}\includegraphics[width=0.25\textwidth]{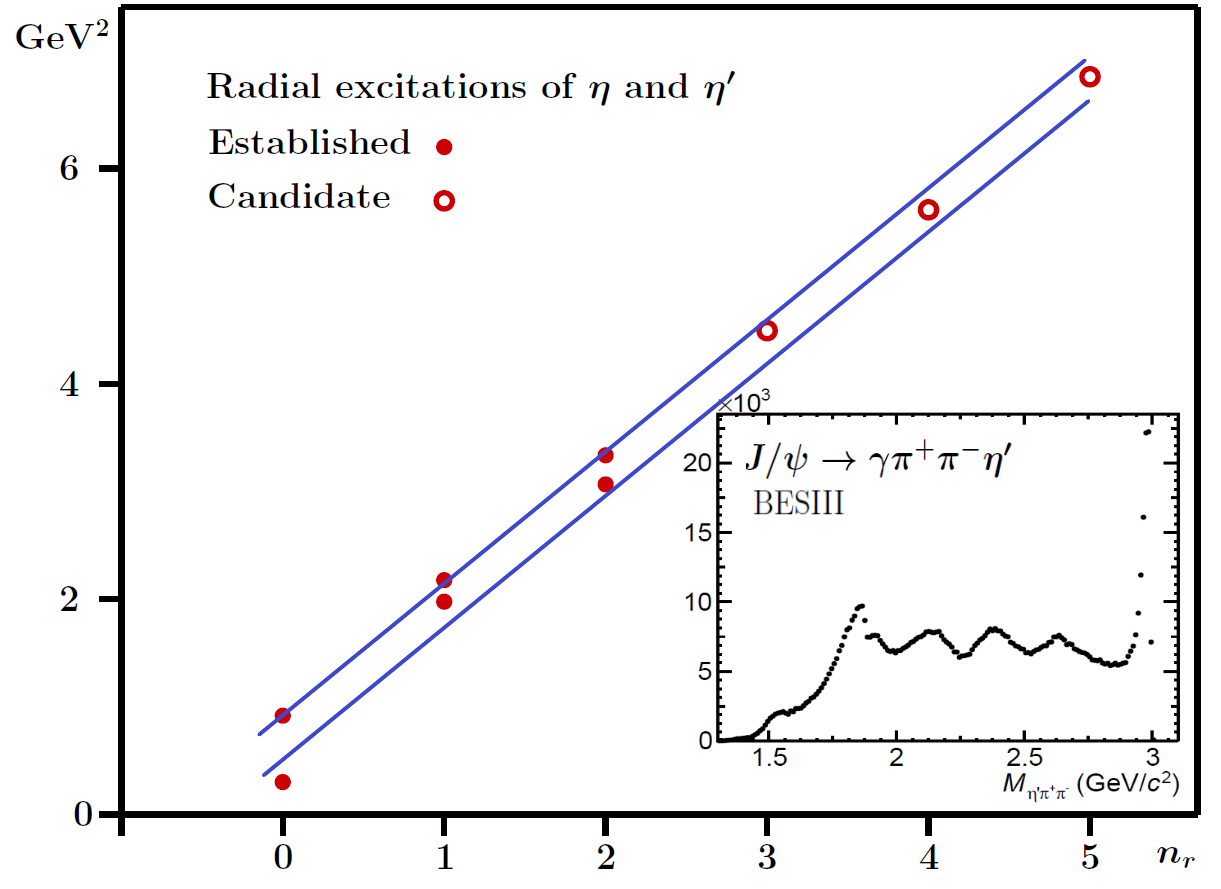}}
\end{tabular}
\end{center}
\caption{\label{PsEK}Left: The $\pi^+\pi^-\eta^\prime$ mass distribution from radiative
$J/\psi$ decays~\cite{BESIIICollaboration:2022kwh}. The quantum numbers are not known. Right: $M^2$ versus n trajectories.
}
\end{figure}
The BESIII collaboration has studied the reaction $J/\psi\to \pi^+\pi^-\eta^\prime$~\cite{BESIIICollaboration:2022kwh}.
The left panel of Fig.~\ref{PsEK}  shows the $\pi^+\pi^-\eta^\prime$  invariant mass distributions with a series of peaks.
Assuming that these are all pseudoscalar mesons, two trajectories can be drawn (right panel of Fig.~\ref{PsEK}).
The figure suggests that the higher-mass structures could house two mesons, possibly singlet and octet
states in SU(3). If this is true, a cut in the $\pi^+\pi^-$ invariant mass at about 1480\,MeV would
partly separate the two isobars, $X(2600)$$\to$$f_0(1370)\eta^\prime$  and $X(2600)$ $\to$$f_0(1500)\eta^\prime$.
We may expect a slight mass shift in the two  $\pi^+\pi^-\eta^\prime$ invariant mass distributions. The
two mesons \,$f_0(1370)$ \,and \,$\eta^\prime$ \,are \,both \,mainly \,singlet. \,The $f_0(1370)\eta^\prime$ isobar
as singlet meson in the $X(2600)$ complex should be slightly higher in mass than the
$f_0(1500)\eta^\prime$
mainly octet meson.

The total yields of the high-mass structures -- including unseen decay modes -- are not known.
Nevertheless, their appearance above a comparatively low background is surprising. Personally,
I suppose that the pseudoscalar glueball is rather wide, and that the structures are seen so clearly
because of a small glueball content. More studies of theses data and of different channels are
required to substantiate this conjecture.

\section{Outlook}
The data of the BESIII collaboration presented above are based on $1.3\cdot 10^9$
events taken at the $J/\psi$. Presently available are $10^{10}$ events.
Based on this large statistics, rare radiative decays like
$J/\psi\to\gamma\eta\eta'$~\cite{BESIII:2022iwi,BESIII:2022riz} and
$J/\psi\to\gamma\eta'\eta'$~\cite{BESIII:2022zel} have been analysed. Data on the different
charge mode of $J/\psi\to\gamma 4\pi$ would be extremely important. In an ideal world,
these data would be publicly available after publication
and would be included in different coupled-channel partial-wave analyses.

\begin{figure}[t]
\centering
\includegraphics[width=0.48\textwidth]{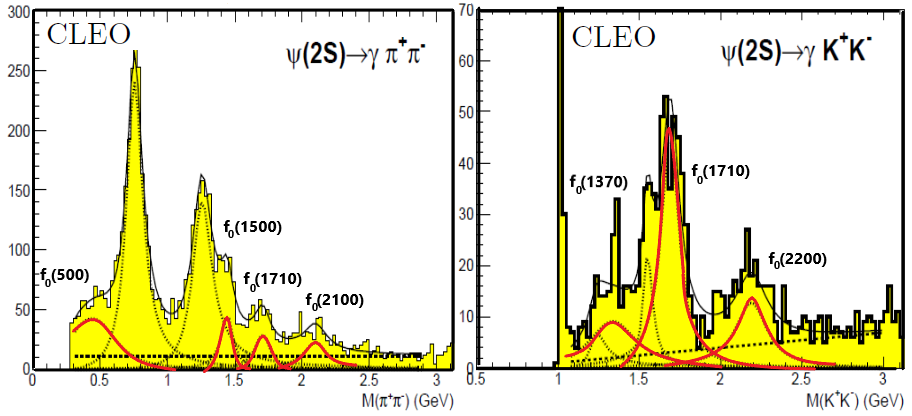}
\caption{\label{fig:psi2-rad}
$\pi^+\pi^-$ (left) and $K^+K^-$ (right) invariant mass distributions from radiative
deacys of $\psi(2S)$. The red curves represent the $S$-wave contributions.
Adapted from~\cite{Dobbs:2015dwa}.\vspace{3mm}
}
\centering
\includegraphics[width=0.48\textwidth]{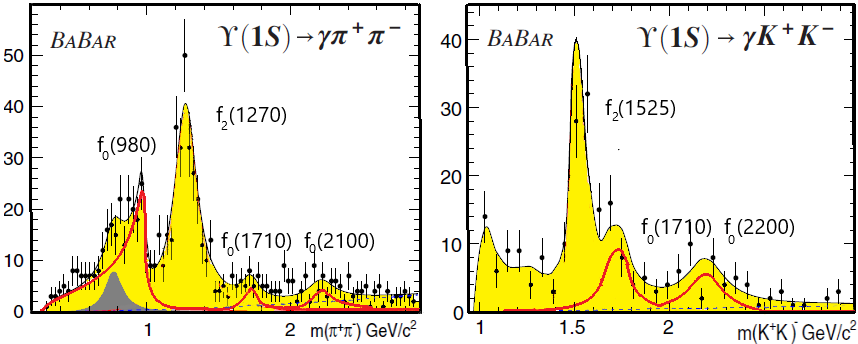}
\caption{\label{fig:Y-rad}
$\pi^+\pi^-$ (left) and $K^+K^-$ (right) invariant mass distributions from radiative
deacys of $\Upsilon(1S)$. The $\Upsilon(1S)$ is observed in $\Upsilon(2S)/\Upsilon(3S)\to\pi^+\pi^-\Upsilon(1S)$. The red curves represent the $S$-wave contributions, the grey area the
$\rho(770)$ contribution. Adapted from~\cite{BaBar:2018uqa}.
}
\end{figure}

Radiative decays of $\psi(2S)$ and of $\Upsilon(1S)$ open a wider range in invariant mass.
The  authors of Ref.~\cite{Dobbs:2015dwa} used the data of the CLEO collaboration on radiative
$\psi(2S)$ decays into $\pi^+\pi^-$ and $K^+K^-$. The data are shown
in Fig.~\ref{fig:psi2-rad}. The data are fit with known resonances,
no partial-wave analysis was performed. The BaBar collaboration studied radiative
$\Upsilon(1S)$ decays into $\pi^+\pi^-$ and
$K^+K^-$~\cite{BaBar:2018uqa}. The results are shown in Fig.~\ref{fig:Y-rad}.
In all four distributions, there is not a single prominent peak in the $S$-wave
contribution which would stick out as glueball candidate. The $S$-waves rather
resembles the distributions observed in radiative $J/\psi$: three major enhancement in the
1500, 1750 and 2200\,MeV region separated by dips. (With the larger statistics in $J/\psi$ decays,
a fourth enhancement is seen at about 2350\,MeV.) In Fig.~\ref{fig:psi2-rad}, a peak
is found at 1447\,MeV and assigned to $f_0(1500)$. At 1500\,MeV, there is the dip. The
wrong mass is due to the neglect of interference: The phase between $f_0(1500)$
and the ``background" (due to the wider $f_0(1370)$) is 180$^\circ$~\cite{Sarantsev:2021ein}. This phase difference
and the significant $f_0(1500)\to\eta\eta'$ branching ratio identify $f_0(1500)$
as mainly SU(3)$_F$ octet state. The different masses for the high-mass state
in the $\pi^+\pi^-$ and $K^+K^-$ invariant mass distributions point
again to the neglect of interference between the prominent octet states and the
singlet ``background". Inspecting Figs.~\ref{fig:psi2-rad} and~\ref{fig:Y-rad}
shows: there is no striking isolated peak which could be interpreted
as ``the glueball''. The glueball content must be distributed over a large number of states.

In $\psi(2S)$ radiative decays,
the $f_0(1710)\to K\bar K$ is observed with a branching fraction of (6.7\er0.9)$\cdot 10^{-5}$,
in $\Upsilon(1S)$ radiative decays, the $f_0(1710)\to K^+ K^-$ is seen with a branching ratio of
(2.02\er0.51\er0.35)$\cdot 10^{-5}$.
The comparison with the yield observed in Ref.~\cite{Sarantsev:2021ein} allows us
to calculated the branching ratio expected for $\psi(2S)$
and $\Upsilon(1S)$ decays when the full
scalar glueball is covered, i.e. for $\Upsilon(1S)\to \gamma G_0(1865)$. The values are given in
Table~\ref{tab:psi2sy1s}.

\begin{table}[t]
\caption{\label{tab:psi2sy1s}Radiative yields expected for $\psi(2S)$
and $\Upsilon (1S)$ radiative decays into the scalar glueball.
}
\renewcommand{\arraystretch}{1.2}
\begin{tabular}{lccc}
\hline\hline
         & ``Exp." & Theory & Ref.\\
$\psi(2S)\to \gamma G_0(1865)$     &\hspace{-2mm}$\sim 5\cdot 10^{-4}$ &  $(5.9^{+3.4}_{-1.4})\cdot 10^{-4}$ & \cite{Zhu:2015qoa}\\
$\Upsilon(1S)\to \gamma G_0(1865)$ &\hspace{-2mm}$\sim 3\cdot 10^{-4}$& $(1.3^{+0.7}_{-0.3})\cdot 10^{-4}$ &
\cite{Zhu:2015qoa}\\
&&$(1 - 2)\cdot 10^{-3}$&\cite{He:2002hr}\\
\hline\hline
\end{tabular}
\end{table}

Clearly, a significant increase in statistics is required when these reactions
should make in independent impact. The advantage of $\psi(2S)$ and $\Upsilon(1S)$
radiative decays is of course that phase space limitations play no role any more. This is particularly important for the search for the tensor and pseudoscalar glueball. The scalar glueball seems to be confirmed: there is not much intensity above
2500\,MeV.

At the end I would like to give an answer to the question posed in the title: yes, I am convinced, the scalar glueball is identified,
and the tensor glueball seems to have left first traces in the data.
\bibliography{Glueballs}
\bibliographystyle{ieeetr}
\end{document}